\documentclass[12pt]{article}
\usepackage{arxiv}

\usepackage[utf8]{inputenc} %
\usepackage[T1]{fontenc}    %
\usepackage{url}            %
\usepackage{booktabs}       %
\usepackage{amsfonts}       %
\usepackage{nicefrac}       %
\usepackage{microtype}      %
\usepackage{xcolor}         %
\usepackage[inline]{enumitem} %
\usepackage{natbib}
\usepackage{doi}
\usepackage{algorithm} %
\usepackage{algpseudocode} %
\algnewcommand\algorithmicinput{\textbf{Input:}}
\algnewcommand\algorithmicoutput{\textbf{Output:}}
\algnewcommand\algorithmicnote{\textbf{Note:}}
\algnewcommand\Input{\item[\algorithmicinput]}%
\algnewcommand\Output{\item[\algorithmicoutput]}%
\algnewcommand\Note{\item[\algorithmicnote]}%

\usepackage{amsmath}
\usepackage{amssymb}
\usepackage{mathtools}
\usepackage{amsthm}
\usepackage{mathrsfs}
\usepackage{multirow, array, threeparttable}
\usepackage[font=normalsize,skip=4pt]{caption}
\usepackage{subcaption}
\usepackage{graphicx}   %
\hypersetup{
  colorlinks,
  linkcolor={blue!60!black},
  citecolor={red!60!black},
  urlcolor={blue!80!black}
}
\usepackage[capitalize,nameinlink]{cleveref}

\makeatletter
\renewcommand{\paragraph}{%
  \@startsection{paragraph}{4}{\z@}%
  {0.25\baselineskip}  %
  {-0.25em}            %
  {\normalfont\normalsize\bfseries}%
}
\makeatother

\usepackage{euscript} %
\usepackage{bm} %

\usepackage{siunitx} %
\usepackage[all, cmtip]{xy}
\usepackage{tikz-cd}
\usepackage[strings]{underscore}
\usepackage{threeparttable} %
\usepackage{multirow} %
\DeclareMathOperator{\diag}{diag}

\title{Nonparametric Stochastic Subspaces via the Bootstrap Method for Characterizing Model Error}

\author{{\hspace{1mm}Akash Yadav} \\
  University of Houston\\
  \texttt{ayadav4@uh.edu} \\
  \And
  {\hspace{1mm}Ruda Zhang} \\
  University of Houston\\
  \texttt{rudaz@uh.edu} \\
}

\date{}

\renewcommand{\headeright}{ }
\renewcommand{\shorttitle}{SS-Bootstrap}

\begin{document}
\maketitle

\begin{abstract} %
Reliable forward uncertainty quantification in engineering requires methods that account for %
aleatory and epistemic uncertainties.
In many applications, epistemic effects arising from uncertain %
parameters and model form dominate prediction error and strongly influence engineering decisions.
Because distinguishing and representing each source separately is often infeasible, their combined effect is typically analyzed using a unified model-error framework. 
Model error directly affects model credibility and predictive reliability; %
yet its characterization remains challenging.
To address this need, we introduce a bootstrap-based stochastic subspace model for characterizing model error in the stochastic reduced-order modeling framework.
Given a snapshot matrix of state vectors,
the method leverages the empirical data distribution to induce a sampling distribution over principal subspaces for reduced order modeling.
The resulting stochastic model enables improved characterization of model error in computational mechanics
compared with existing approaches.
The method offers several advantages:
(1) it is assumption-free and leverages the empirical data distribution;
(2) it enforces linear constraints (such as boundary conditions) by construction;
(3) it requires only one hyperparameter, significantly simplifying the training process; and
(4) its algorithm is straightforward to implement.
We evaluate the method's performance against existing approaches using numerical examples in computational mechanics and structural dynamics.
\end{abstract}

\keywords{Model error \and Model-form uncertainty \and Stochastic reduced-order modeling \and Bootstrap \and Non-parametric method}

\section*{Practical Applications}
Engineers often rely on computational models for simulation, design, and other engineering decisions.
However, computational models usually contain errors from various sources and can often be computationally expensive to run. 
Projection-based reduced-order models are an efficient way to reduce cost but also introduce additional mode truncation errors.
The proposed method, SS-Bootstrap, builds a stochastic reduced-order model that not only provides fast predictions but also estimates model uncertainty.
The approach is relevant to applications such as structural dynamics,
digital twins, model validation, design exploration,
and reliability assessment,
where repeated high-fidelity simulations are too expensive for routine use.

A key benefit of the method is that it requires only snapshot data from an existing computational model and introduces just one tunable hyperparameter, making it relatively easy to train and implement.
It can be used both to quantify error caused by model reduction
and to represent discrepancies between computational predictions and observed data.
In the examples studied here,
the method produced uncertainty bounds that were both statistically reliable and sharper than those from an earlier stochastic-subspace approach.
This makes the framework attractive for practical workflows where repeated high-fidelity simulations are too expensive, but decision-makers still need trustworthy error estimates.

\section{Introduction}

Engineering systems often exhibit uncertainty,
which may arise from various sources,
including the intrinsic variability of the system and incomplete knowledge of the governing physics.
In the context of uncertainty quantification (UQ),
these are commonly categorized into aleatory and epistemic uncertainties.
\textit{Aleatory} uncertainty stems from intrinsic variability in systems and data,
such as fluctuating boundary conditions, inherently variable material properties,
environmental conditions, or measurement noise,
and is often irreducible \citep{Kiureghian2009}.
\textit{Epistemic} uncertainty results from incomplete knowledge,
modeling assumptions, and numerical approximations;
unlike aleatory uncertainty, it can in principle be reduced with deeper physical understanding, improved models, or additional data \citep{Oberkampf2002,Roy2011}.

Within epistemic uncertainty, multiple sources such as uncertain inputs, 
parameters, and model form typically coexist.
Errors in model parameters and model form are often intertwined; a practical and widely used perspective is to analyze their combined effect through a unified \textit{model-error} framework.
Model error refers to the discrepancy between model predictions and ground truth;
in various contexts, it is also called 
model inadequacy/model discrepancy \citep{Kennedy2001},
or structural uncertainty \citep{Trucano2006}.
It arises when the numerical model fails to accurately capture the behavior of the true system,
resulting in a systematic prediction error (i.e., model error) that extends beyond parametric misspecification.
Treating model-form uncertainty as parametric uncertainty can lead to biased and overconfident predictions \citep{Arendt2012,Tuo2016,Pernot2017}. 
Model error is often a key factor in epistemic uncertainty, but it is a challenging problem.
Model error analysis is typically divided into two components:
characterization and correction.
\textit{Characterization} refers to quantifying the predictive error of an inadequate model,
whereas \textit{correction} involves reducing the predictive error of an inadequate model via some form of adjustment.
Epistemic uncertainty may be described using non-probabilistic frameworks, including interval \cite{Chong2018} and fuzzy models \cite{HANSS2010}. 
More often, works on this topic provide probabilistic representations of model error,
which can be categorized into parametric and nonparametric approaches.

\textit{Parametric approaches} employ a statistical model that specifies a family of distributions with a fixed, finite number of parameters.
In model-error analysis, such parameterizations
are attractive because they typically yield tractable inference.
Representative examples include stochastic closure models in governing equations,
and parametric representations of discrepancy in numerical models and surrogate models.
\citet{Xiao2016} present a data-driven, physics-informed Bayesian framework based on an iterative ensemble Kalman method to assimilate prior knowledge and observation data to quantify and reduce model-form uncertainties in Reynolds-averaged Navier-Stokes simulations.
\citet{Morrison2018} address model inadequacy with stochastic operators
introduced as a source in the governing equations while preserving underlying physical constraints.
Similarly, \citet{Portone2022} propose a stochastic operator
for an unclosed dispersion term in an averaged advection--diffusion equation
for model closure.
\citet{Sargsyan2019EmbeddedModelError} provide a Bayesian framework with embedded model error using polynomial chaos expansions (PCEs)
for improved uncertainty representation in computational models. 
\citet{WangZH2023} introduce a Bayesian parametric model for statistical calibration of model error induced by uncertainty in PCE coefficients.
\citet{Zou2024} present a method for correcting model misspecification in physics-informed neural networks (PINNs) using Bayesian PINNs and ensemble PINNs.
Parametric model-error formulations span a broad literature,
and these works are intended as an illustrative sample.

\textit{Nonparametric approaches} do not restrict the data-generating mechanism to a fixed finite-dimensional form:
the unknown object is either infinite-dimensional or its effective dimension grows with the data.
Nonparametric methods are flexible and robust to model misspecification, but they can be challenging to use for inference and computation as data sizes increase.
Consequently, relatively few methods fall into this category.
One of the earliest nonparametric approaches for model error is the Kennedy–O’Hagan (KOH) framework, which corrects model error by learning a Gaussian process (GP) discrepancy from data, providing a flexible representation without prescribing a fixed form \citep{Kennedy2001}. 
Despite its wide adoption \citep{Higdon2004,Higdon2008,Qian2008}, the KOH framework can exhibit identifiability issues between calibration parameters and discrepancy and may extrapolate unreliably without informative priors or physics-based structure \citep{Strong2014,Brynjarsdottir2014,Sargsyan2015,Plumlee2017,Maupin2020}.
Beyond output-discrepancy models, some nonparametric strategies place GPs directly on operators and fields.
For example, \citet{Girolami2021} model the forcing term and the coefficients of the differential operator as Gaussian processes and develop a statistical finite element method that coherently synthesizes observation data and model predictions.

A recent nonparametric model (NPM) for model uncertainty \citep{Soize2017srob} merges projection-based reduced-order modeling %
with random matrix theory. %
This framework uses a randomized reduced-order basis (ROB)
to induce a stochastic reduced-order model (ROM),
which allows efficient characterization of model error
as it involves computationally inexpensive ROMs instead of full-order models.
The stochastic ROBs used in the NPM are defined through a sampling procedure:
samples from a Gaussian process are stacked into a matrix, and boundary conditions are explicitly enforced;
the matrix is then used to construct a perturbed version of the deterministic ROB through tangential projection and polar decomposition.
Hyperparameters of the stochastic ROB are identified by solving a reduced statistical inverse problem.
The NPM method has been applied to various engineering problems \citep{WangHR2019,Soize2019srob},
with recent efforts focused on reducing the number of hyperparameters and simplifying training \citep{Soize2019srob,Azzi2022}.

Following the stochastic ROM framework but designed for multi-model settings,
the Riemannian stochastic model (RSM) \citep{ZhangH2023} works with a different set of stochastic bases:
a Dirichlet distribution anchored to the candidate models' ROBs is constructed on a tangent space of the Stiefel manifold,
and its samples are reconstructed as bases via the Riemannian exponential map.
The RSM estimates its hyperparameters by solving a quadratic program and has been applied in computational mechanics \citep{Zhang2024, Quek2025}.

In prior work, we introduced SS-PPCA, which models principal subspaces via probabilistic PCA \citep{yadav2025ss}. 
Also based on a stochastic ROM framework, SS-PPCA differs from NPM and RSM in its modeling approaches:
NPM and RSM model stochastic ROBs, whereas SS-PPCA models the subspaces directly.
SS-PPCA provides analytical distributions for the associated random matrices and subspaces
and involves only a single hyperparameter, which simplifies both its implementation and training.
However, it relies on a Gaussian latent variable model, which, while often effective,
may not perform well for heavy-tailed, skewed, or multimodal data.

In this paper, we present a nonparametric, data-driven framework—stochastic subspace via bootstrap (SS-Bootstrap)—for probabilistic modeling of principal subspaces and characterization of model error.
Unlike SS-PPCA, which relies on a Gaussian latent-variable structure, SS-Bootstrap operates directly on the empirical data distribution via resampling, leveraging classical bootstrap principles \citep{EfronBootstrap1979,Efron1994}. 
This makes the framework broadly applicable, assumption-light, and naturally robust to non-Gaussian and multimodal data. 
The main contributions of our work are as follows:
\begin{enumerate}
    \item We propose a novel non-parametric, data-driven framework for stochastic subspace modeling based on bootstrap resampling, which is simple, distribution-free, and computationally efficient.
    \item We use this stochastic subspace model to construct stochastic ROMs with a single scalar hyperparameter. This contrasts with existing stochastic ROM approaches, which use stochastic bases and typically involve more hyperparameters.
    \item We demonstrate, through a series of numerical experiments, that our method provides consistent, sharp uncertainty estimates at low computational cost.
\end{enumerate}

The remainder of the paper is organized as follows. \Cref{sec:SROM} reviews stochastic reduced-order modeling.
\Cref{sec:SS-model} presents the proposed stochastic subspace model, sampling algorithm, and hyperparameter optimization strategy. 
\Cref{sec:related} reviews related work. 
\Cref{sec:examples} validates the proposed method through numerical examples. 
Finally, \cref{sec:conclusion} concludes with a summary and directions for future work.

\textit{Notation.}
Throughout this paper, vectors are denoted by bold lowercase letters (e.g., $\mathbf{x}$),
matrices by bold uppercase letters (e.g., $\mathbf{X}$),
and scalars by regular typeface (e.g., $x_i$).
Subspaces are denoted by calligraphic uppercase letters (e.g., $\mathscr{V}$).
Random quantities share the same notation as their deterministic counterparts;
their nature should be clear from the context.

\section{Stochastic reduced-order modeling}\label{sec:SROM}
This section develops the formulation of stochastic reduced-order models (SROMs).
We first define the Full-order model (FOM), then construct a deterministic reduced-order model (ROM) from the FOM, and finally formulate an SROM based on the ROM. 
The presentation is kept general so that the framework applies across a broad range of physical and engineering systems.

\subsection{Full-order model}\label{sec:HDM}

Consider a parametric nonlinear dynamical system governed by a set of ordinary differential equations (ODEs):
\begin{equation}\label{eq:HDM}
  \dot{\mathbf{x}} = \mathbf{f}(\mathbf{x}, t; \boldsymbol{\mu}),
\end{equation}
where $\mathbf{x} \in \mathbb{R}^n$ is the state vector, $t \in [0, \infty)$ is time, and $\boldsymbol{\mu}$ denotes the parameter vector.
The initial condition is $\mathbf{x}(0) = \mathbf{x}_0$, and the system is subject to linear constraints of the form $\mathbf{B}^\intercal \mathbf{x} = 0$, with $\mathbf{B} \in \mathbb{R}^{n \times n_{CD}}$. 
These constraints often encode boundary conditions.
Nonhomogeneous linear constraints can be recast as homogeneous ones by a state shift, without loss of generality.
Such systems typically arise from the spatial discretization of partial differential equations governing physical processes, resulting in \textit{Full-order models} (FOMs) with $n \gg 1$.
For steady-state problems, \cref{eq:HDM} reduces to an algebraic system $\mathbf{f}(\mathbf{x}; \boldsymbol{\mu}) = 0$.
The FOM is an approximation of the physical system due to modeling and numerical assumptions; the resulting discrepancy is the target for model-error analysis.

\subsection{Reduced-order model}\label{sec:ROM}

Reduced-order models (ROMs) can be constructed from the FOM to accelerate simulations while maintaining accuracy.
A standard approach for ROM construction is the Galerkin projection,
which projects the FOM orthogonally onto a $k$-dimensional subspace $\mathscr{V}$
of the state space $\mathbb{R}^n$.
This reduced subspace $\mathscr{V}$ can be obtained, for example,
using proper orthogonal decomposition (POD) \citep{Sirovich1987} applied to a snapshot matrix of FOM solutions, or from the eigendecomposition of the system operators \citep{Chopra1996}.
For POD, the reduced order $k$ is typically selected to meet a prescribed energy threshold retained by the POD modes.
Let $\mathbf{V} \in \text{St}(n, k)$ be an orthonormal basis of the subspace $\mathscr{V}$ where $\text{St}(n, k)$ denotes Stiefel manifold,
then the ROM can be written as:
\begin{equation}\label{eq:ROM}
  \mathbf{x} = \mathbf{V} \mathbf{q},
  \quad
  \dot{\mathbf{q}} = \mathbf{V}^\intercal \, \mathbf{f}(\mathbf{V} \mathbf{q}, t; \boldsymbol{\mu}),
\end{equation}
with reduced coordinates $\mathbf{q} \in \mathbb{R}^k$
and initial conditions $\mathbf{q}(0) = \mathbf{V}^\intercal \mathbf{x}_0$.
To satisfy the linear constraints, we must have $\mathbf{B}^\intercal \mathbf{V}= 0$. 
Since POD bases are constructed from snapshot data, the resulting ROM automatically satisfies any linear constraints (homogeneous or non-homogeneous) present in the snapshots. 
In steady-state setting, \cref{eq:ROM} reduces to a set of algebraic equations:
$\mathbf{V}^\intercal \, \mathbf{f}(\mathbf{V} \mathbf{q}; \boldsymbol{\mu})$ = $0$. 
Although ROMs offer substantial computational savings, they introduce additional approximation error, particularly due to the mode truncation. 
These errors contribute to model-form uncertainty and must be analyzed.

When the FOM depends on parameters,
it is often beneficial for the reduced subspace to vary with them,
leading to a parameter-to-subspace map $\mathscr{V}(\boldsymbol{\mu})$. 
One can learn such maps from data via interpolation or regression.
For example, the Gaussian process subspace (GPS) model places a Gaussian process prior
on an extrinsic representation of the basis, inducing a predictive distribution over
$k$-dimensional subspaces across the parameter domain \citep{ZhangRD2022gps}. 
Allowing $\mathscr{V}(\boldsymbol{\mu})$ to vary typically improves accuracy over using a fixed $k$-dimensional subspace.
However, to keep the exposition focused, we restrict attention to a fixed subspace in this paper.

\subsection{Stochastic reduced-order model}\label{sec:sub-SROM}

A stochastic reduced-order model (SROM) extends the deterministic ROM by treating the reduced basis as a random quantity. 
Specifically, instead of the deterministic basis $\mathbf{V}$, the SROM uses a stochastic basis $\mathbf{W}$.
The change from a deterministic basis to a stochastic one
introduces randomness into the model and allows it to capture
model errors between the ROM, the FOM, and the physical system.
The SROM can be written as:
\begin{equation}\label{eq:SROM}
  \mathbf{x} = \mathbf{W} \mathbf{q},
  \quad \dot{\mathbf{q}} = \mathbf{W}^\intercal \, \mathbf{f}(\mathbf{W} \mathbf{q}, t; \boldsymbol{\mu}),
  \quad \mathbf{W} \sim \mu_{\mathbf{V}}
\end{equation}
where stochastic basis $\mathbf{W}$ follows a probability distribution $\mu_{\mathbf{V}}$.
However, it is more natural to impose a probability distribution $\mu_{\mathscr{V}}$ on the subspace rather than the basis,
because the Galerkin projection is uniquely determined by the subspace and is invariant under changes of basis.
In this work, we use the stochastic subspace model introduced in \cref{sec:SS-model}
to sample stochastic basis $\mathbf{W}$ in a distribution-free and data-driven manner. 
These stochastic bases form the foundation of SROMs, facilitating the effective characterization of model error.

\section{Stochastic subspace model} \label{sec:SS-model}
This section develops a bootstrap-based stochastic subspace model.
We begin by constructing a deterministic principal subspace via proper orthogonal decomposition (POD), then introduce distribution-free modeling with the bootstrap.
Building on these ingredients, we formulate the stochastic subspace model and its associated sampling algorithm.
Finally, we present a hyperparameter optimization strategy and outline the key steps for constructing SROMs.

\subsection{Subspace from the PCA: proper orthogonal decomposition} \label{sec:subspace-PCA}

Principal component analysis (PCA) is widely used for dimension reduction.
Closely related to PCA, \textit{proper orthogonal decomposition} (POD) provides a framework for computing the ROB $\mathbf{V}$ and its associated subspace $\mathscr{V}$.
Following the notations in \cref{sec:SROM},
consider a snapshot matrix $\mathbf{X} = [\mathbf{x}_1 \, \cdots \, \mathbf{x}_m] \in \mathbb{R}^{n \times m}$, with states $\mathbf{x}_i = \mathbf{x}(t_i; \boldsymbol{\mu}_i)$,
sample mean $\overline{\mathbf{x}} = \frac{1}{m} \sum_{i=1}^m \mathbf{x}_i$,
centered snapshot matrix $\mathbf{X}_0 = \mathbf{X} - \overline{\mathbf{x}} \mathbf{1}_m^\intercal$,
and sample covariance $\mathbf{S} = \frac{1}{m} \mathbf{X}_0 \mathbf{X}_0^\intercal$.
Let $\mathbf{X}_0 = \mathbf{V}_r \diag(\boldsymbol{\sigma}_r) \mathbf{W}_r^\intercal$
be a compact singular value decomposition (SVD), where
$\boldsymbol{\sigma}_r \in \mathbb{R}^r_{>0 \downarrow}$ is in non-increasing order.
The sample covariance matrix $\mathbf{S}$ can be written as
$\mathbf{S} = \mathbf{V}_r \diag(\boldsymbol{\sigma}_r / \sqrt{m})^2 \mathbf{V}_r^\intercal$.
POD selects the \textit{principal basis} $\mathbf{V}_k$---the leading $k$ eigenvectors of $\mathbf{S}$---as the deterministic ROB.
The corresponding subspace is the \textit{principal subspace} $\mathscr{V}_k := \text{range}(\mathbf{V}_k)$.
Since all state samples satisfy the linear constraints,
$\mathbf{V}_k$ satisfies the constraints automatically.
POD has several other desirable properties: it can extract coherent structures, provide error estimates in $L^2$ norm,
and is computationally efficient and straightforward to implement \citep{Sirovich1987}.

\subsection{Bootstrap: distribution-free modeling} \label{sec:modeling-bootstrap}

The bootstrap \citep{EfronBootstrap1979} is a nonparametric method that approximates the unknown sampling distribution of a statistic by resampling from the data's empirical distribution. 
It avoids parametric assumptions about the population distribution and is effective when classical assumptions (e.g., Gaussianity) may be violated.

Given a random sample $\mathbf{X}=[\mathbf{x}_1 \, \cdots \, \mathbf{x}_m]$ drawn from an unknown distribution $\mathcal{F}$,
the bootstrap method can approximate the sampling distribution of some prespecified random variable $r(\mathbf{X}; \mathcal{F})$
using the observed data. %
The bootstrap method constructs the sample probability distribution $\hat{\mathcal{F}}$ by putting mass $1/m$ at each observation point $\mathbf{x}_1, \cdots, \mathbf{x}_m$.
Sampling from the empirical distribution $\hat{\mathcal{F}}$ can be performed by selecting indices randomly:
\begin{equation*}
    \mathbf{X}^\ast = [\mathbf{x}_{b_1} \, \cdots \, \mathbf{x}_{b_m}], \quad b_i \overset{\text{iid}}{\sim} U(\{1, \, \cdots, \, m\}), \ i \in \{1, \cdots, m\}.
\end{equation*}
Here, $U(\{1, \, \cdots, \, m\})$ refers to the uniform distribution over the index set $\{1, \, \cdots, \, m\}$.
In simple terms, the bootstrap sample $\mathbf{X}^\ast$ is obtained by sampling the observed data with replacement.
The bootstrap samples are then used to approximate the sampling distribution of $r(\mathbf{X}; \mathcal{F}) $
by that of $r^\ast = r(\mathbf{X}^\ast; \hat{\mathcal{F}})$.
When the observed data follow a Gaussian distribution, bootstrap samples can recover that behavior. 
The main advantage of the nonparametric bootstrap emerges when these assumptions are violated: for skewed, heavy-tailed, or multimodal data, resampling from $\hat{\mathcal{F}}$ propagates the empirical features of the data—such as asymmetry, tail behavior, and multimodality—into the distribution of $r^\ast$.
In this way, the bootstrap can reveal uncertainty structures that Gaussian assumptions would obscure, while remaining fully data-driven and free of an explicit parametric model for $\mathcal{F}$.

Bootstrap methods have been extended beyond Euclidean settings
to respect intrinsic geometric structures.
In particular, the normal-bundle bootstrap constructs resamples that preserve a learned data manifold
while randomizing along its normal directions \citep{ZhangRD2021nbb}.
The same geometric perspective applies to our setting:
in SROMs the randomized objects are $k$-dimensional subspaces, which are points on the Grassmann manifold.
Probabilistic models for such subspaces should therefore preserve this manifold structure.
Previous works use the matrix angular central Gaussian family, a parametric class of distributions on the Grassmannian,
to describe random subspaces \citep{ZhangRD2022gps,yadav2025ss}.
Preserving the geometric constraint while increasing modeling flexibility
motivates the bootstrap-based stochastic subspace model introduced next.

\subsection{Stochastic subspace via the bootstrap} \label{sec:stochastic-subspace-bootstrap}

We now propose a bootstrap-based stochastic subspace model.
Here, the prespecified random variable is the principal subspace, and we approximate its distribution using the bootstrap resampling method. 

Consider the \textit{principal subspace map} $\pi_k(\mathbf{X}) := \text{range}(\mathbf{U}_k)$,
where $\mathbf{U}_k$ consists of left singular vectors associated with
the $k$ largest singular values of $\mathbf{X}$.
As a mapping, $\pi_k: \mathbb{R}^{n \times m}_{k>} \mapsto \text{Gr}(n, k)$ is uniquely defined
for all $n$-by-$m$ matrices whose $k$-th largest singular value is larger than the $(k+1)$-th,
with $k \le \min(n, m)$.
Using this map, the POD subspace can be written as:
\begin{equation}
    \mathscr{V}_k = \pi_k(\mathbf{X}_0).
\end{equation}
In classical bootstrap methods, each replicate contains the same number of samples as the observed dataset.
Here, we introduce a hyperparameter $\beta$ that decouples the bootstrap sample size from the data size,
thereby controlling the effective sample size of the resampled snapshot matrix.

Let $\mathbf{X}^\ast_{0,\beta}$ denote a bootstrap-resampled snapshot matrix
obtained by sampling the columns of $\mathbf{X}_0$ randomly with replacement for $\beta$ times and concatenating them into a matrix.
Applying the principal subspace map to $\mathbf{X}^\ast_{0,\beta}$ yields a random principal subspace,
and repeating this procedure generates a collection of stochastic subspaces.
Formally, we define the stochastic subspace model:
\begin{equation}\label{eq:dist-nkp}
  \mathscr{W} := \pi_k(\mathbf{X}^\ast_{0,\beta}),
\end{equation}
where $\mathbf{X}^\ast_{0,\beta} \sim \hat{\mathcal{F}}$
is a bootstrap-resampled snapshot matrix with $\beta \in \{k, k+1, \cdots\}$ columns.
The close connection of this model with the POD justifies naming it as a \textit{stochastic POD}.
The scalar hyperparameter $\beta$ controls the concentration of the distribution:
a larger value corresponds to less variation around $\mathscr{V}_k$. 
In particular, smaller $\beta$ leads to greater dispersion in the stochastic subspace distribution, while larger $\beta$ concentrates the distribution more tightly around the deterministic ROM subspace; in the limit of very large $\beta$, the SROM prediction approaches the deterministic ROM prediction.
Due to its intrinsic POD connection, any stochastic basis $\mathbf{W}$ sampled from \cref{eq:dist-nkp} naturally satisfies linear constraints, i.e., $\mathbf{B}^\intercal \mathbf{W} = 0$.

A naive implementation of \cref{eq:dist-nkp} requires a bootstrap resampling of $\mathbf{X}_{0}$
and a truncated SVD, which can be costly when $n$ is large.
However, sampling from this stochastic subspace model
can be done very efficiently for $n \gg 1$
if $k$ and the rank of $\mathbf{X}_0$ are small,
which is often the case in practical applications.
Let $\mathbf{X}_0 = \mathbf{V}_r \diag(\boldsymbol{\lambda}_r) \mathbf{W}_r^\intercal$
be a compact SVD where $r = \text{rank}(\mathbf{X}_0)$.
We can show that:
\begin{equation}\label{eq:dist-nkp-low-rank}
  \mathscr{W} = \mathbf{V}_r\, \pi_k\left(\diag(\boldsymbol{\lambda}_r)
    \mathbf{W}_r(\mathbf{b},:)^\intercal \right),
\end{equation}
where $\mathbf{b} = (b_i)_{i = 1}^\beta$ with entries
$b_{i} \overset{\text{iid}}{\sim} U(\{1, \cdots, m\})$.
\Cref{alg:ss-Bootstrap} describes the corresponding sampling procedure.
Using \cref{eq:dist-nkp-low-rank} instead of \cref{eq:dist-nkp}
reduces the computational cost for truncated SVD from $O(n k \beta)$ to $O(r k \beta)$.
Consequently, when applied to the SROM framework, the bootstrap-based stochastic subspace model offers a robust, efficient, and distribution-free mechanism for characterizing and quantifying model-form uncertainty in computational mechanics.

\alglanguage{pseudocode}
\begin{algorithm}[!t]
  \caption{\texttt{SS-Bootstrap}: Stochastic subspace via Bootstrap.}
  \label{alg:ss-Bootstrap}
  \begin{algorithmic}[1] %
    \Input $\mathbf{V}_r; \diag(\boldsymbol{\lambda}_r); \mathbf{W}_r$;
    subspace dimension $k \in \{1, \cdots, n\}$;
    resample size $\beta \in \{k, k+1, \cdots\}$. %
    \State Generate vector $\mathbf{b} = (b_i)_{i = 1}^\beta$ with entries
    $b_{i} \overset{\text{iid}}{\sim} U(\{1, \cdots, m\})$.
    \State $\mathbf{M} \gets \diag(\boldsymbol{\lambda}_r) \mathbf{W}_r(\textbf{b},:)^{\intercal}$,
    where (\textbf{b},:) refers to the row selection of matrix $\mathbf{W}_r$.
    \State Truncated SVD: $[\mathbf{U}_k, \sim, \sim] \gets \text{svd}(\mathbf{M}, k)$.
    \Output $\mathbf{W} = \mathbf{V}_r \mathbf{U}_k$,
    an orthonormal basis of a random subspace sampled from the bootstrap model.
  \end{algorithmic}
\end{algorithm}

\subsection{Hyperparameter training}
\label{sec:training}

The hyperparameter $\beta \in \{k,k{+}1,\dots\}$ is estimated by minimizing the objective function
\begin{equation} \label{eq:objective}
  f(\beta) := \mathbb{E}[|d_o(\mathbf{u}_L) - d_o(\mathbf{u}_E)|^2 \, | \, \beta],
\end{equation}
where $\mathbf{u}_E$ is the experimental or ground-truth observation of the output,
$\mathbf{u}_L$ is the low-fidelity prediction of the SROM,
and $d_o(\mathbf{u}) := \|\mathbf{u} - \mathbf{u}_L^o\|_{L^2}$ is the $L^2$ distance
to the low-fidelity prediction $\mathbf{u}_L^o$ of a reference model.
This objective function aims to improve the consistency of the SROM
in characterizing the error of the reference model.
We optimize the objective function $f(\beta)$ efficiently using our Bayesian optimization under uncertainty framework tailored to concentration-parameter training in stochastic models \citep{Yadav2025SO-BO-scale}. 
The Bayesian optimization under uncertainty framework uses a statistical surrogate for the underlying random variable, enabling the analytical evaluation of the expectation operator. 
Moreover, it provides a closed-form expression for the optimizer of the random acquisition function, which significantly reduces the computational cost per iteration compared to optimization approaches based on Monte Carlo approximation of the expectation. 

\subsection{Construction of SROM using SS-Bootstrap}

\begin{figure}[!t]
    \centering
    \includegraphics[width=0.5\textwidth]{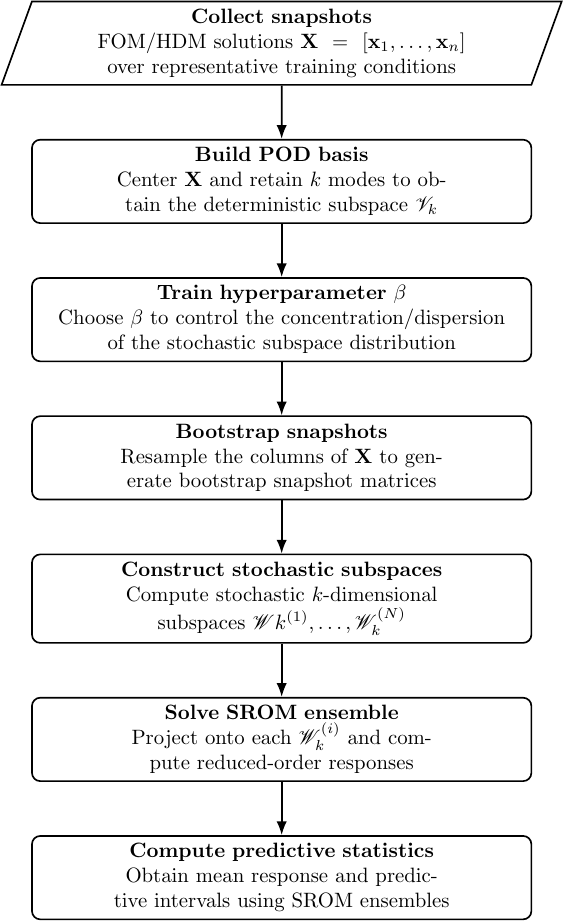}
    \caption{Flowchart of the proposed SS-Bootstrap method.}
    \label{fig:flowchart}
\end{figure}

The proposed SS-Bootstrap approach to stochastic reduced-order modeling begins by solving the FOM and collecting state snapshots at selected times and/or parameter values to form the snapshot matrix $\mathbf{X} = [\mathbf{x}_1,\ldots,\mathbf{x}_m]$.
The snapshot set should be chosen so that the regime of interest is adequately represented; in the parametric setting, this means sampling the parameter domain according to the prescribed input distribution.
The snapshots are then centered as $\mathbf{X}_0 = \mathbf{X} - \overline{\mathbf{x}}\mathbf{1}_m^\intercal$, where $\overline{\mathbf{x}} = \frac{1}{m}\sum_{i=1}^m \mathbf{x}_i$, and a compact singular value decomposition is computed, $\mathbf{X}_0 = \mathbf{V}_r \diag(\boldsymbol{\sigma}_r)\mathbf{W}_r^\intercal$.
The deterministic reduced-order basis is obtained by selecting $k$ as the smallest integer $j$ such that
$\sum_{i=1}^{j}\sigma_i^2 \ge \tau \sum_{i=1}^{r}\sigma_i^2$,
where $\tau \in (0,1)$ is a prescribed variance threshold, and then taking the first $k$ singular vectors $\mathbf{V}_k$.
Substituting $\mathbf{V}_k$ into \cref{eq:ROM} gives the deterministic ROM. To construct the stochastic extension, SS-Bootstrap is used; see \cref{alg:ss-Bootstrap}.
In particular, stochastic subspaces $\mathscr{W}$ with bases $\mathbf{W}=\mathbf{V}_r\mathbf{U}_k$ are generated, and the concentration parameter $\beta \in [k,\infty)$ is tuned according to the criterion in \cref{eq:objective}.
Once $\beta$ is selected, samples of $\mathscr{W}$ are used to generate stochastic bases $\mathbf{W}$ that replace $\mathbf{V}_k$ in \cref{eq:ROM}, yielding an ensemble of SROM realizations whose variability quantifies model and truncation errors (see \cref{fig:flowchart} for more details).

These steps are applicable to both linear and nonlinear systems. For linear systems, however, the cost of SROM construction can be substantially reduced through a two-stage approach.
A straightforward implementation would require, for each SROM realization, constructing a stochastic basis $\mathbf{W}=\mathbf{V}_r\mathbf{U}_k$ and projecting the FOM operators onto that basis, which becomes expensive as the number of samples increases.
A more efficient strategy is to first project the operators once onto the rank-$r$ subspace spanned by $\mathbf{V}_r$ (for example, $\mathbf{A}_r=\mathbf{V}_r^\intercal \mathbf{A}\mathbf{V}_r$ for any system matrix $\mathbf{A}$), and then, for each sampled $\mathbf{U}_k$, form $\mathbf{A}_{\mathbf{W}}=\mathbf{U}_k^\intercal \mathbf{A}_r \mathbf{U}_k$, where the overall basis is $\mathbf{W}=\mathbf{V}_r\mathbf{U}_k$.
This avoids repeated high-dimensional operator projections and significantly lowers the cost of constructing the SROM ensemble.
The proposed SS-Bootstrap construction is not limited to linear systems, since it depends only on snapshot data and the associated POD subspace.
However, the efficient two-stage construction used here relies on a linear operator structure.
For nonlinear problems, the stochastic subspace can still be constructed in the same way, but efficient online evaluation of the resulting SROM generally requires additional hyper-reduction techniques.
In such settings, there are often two reduced bases involved: one for the state and another for the nonlinear term or residual.
Modeling stochastic versions of these two subspaces independently may not be appropriate, since their coupling can be important for accuracy and stability.

\section{Related work} \label{sec:related}

While NPM and RSM also fall within the SROM framework, both methods model the reduced-order basis, which makes them less suitable for direct comparison with the proposed approach. Consequently, this section reviews SS-PPCA, compares its operational mechanics with those of SS-Bootstrap, and presents evaluation metrics for stochastic models.

\begin{figure}[!t]
    \centering
    \includegraphics[width=0.8\textwidth]{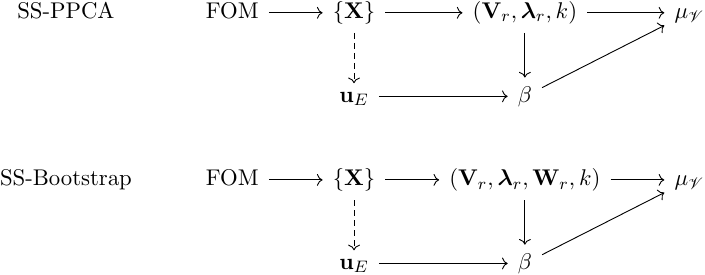}
    \caption{Dependence diagrams for (from top to bottom) the SS-PPCA and the SS-Bootstrap models of stochastic subspace.
    The dashed lines indicate that the connection only exists
    when characterizing the ROM-to-FOM error.}
    \label{fig:comparision}
\end{figure}

\subsection{Stochastic subspace via probabilistic principal component analysis (SS-PPCA)} \label{sec:SS-PPCA}

Given $\mathbf{X} = [\mathbf{x}_1 \, \cdots \, \mathbf{x}_m] \in \mathbb{R}^{n \times m}$ a sample of the state, with $\mathbf{x}_i = \mathbf{x}(t_i; \boldsymbol{\mu}_i)$,
sample mean $\overline{\mathbf{x}} = \frac{1}{m} \sum_{i=1}^m \mathbf{x}_i$,
centered sample $\mathbf{X}_0 = \mathbf{X} - \overline{\mathbf{x}} \mathbf{1}_m^\intercal$,
and sample covariance $\mathbf{S} = \frac{1}{m} \mathbf{X}_0 \mathbf{X}_0^\intercal$, SS-PPCA \citep{yadav2025ss} models the unknown data distribution as $\widetilde{\mathbf{x}} \sim N_n(\overline{\mathbf{x}}, \mathbf{S})$.
The stochastic subspace via SS-PPCA is then defined as:
\begin{equation}\label{eq:MACG-nkp}
  \mathscr{W} := \pi_k(\widetilde{\mathbf{X}}_{0,\beta})
  \overset{\text{def}}{\sim} \text{MACG}_{n,k,\beta}(\mathbf{S}),
\end{equation}
where $\widetilde{\mathbf{X}}_{0,\beta} \sim N_{n,\beta}(0; \mathbf{S}, \mathbf{I}_{\beta})$
is a $\beta$-sample matrix and $\beta \in \{k, k+1, \cdots\}$ is the resample size.

Intuitively, SS-PPCA first postulates a multivariate Gaussian model for the snapshots
with mean $\overline{\mathbf{x}}$ and covariance $\mathbf{S}$,
then draws a random snapshot matrix from this distribution to obtain $\widetilde{\mathbf{X}}_{0,\beta}$.
Applying the map $\pi_k(\cdot)$ extracts the leading $k$-dimensional principal subspace,
inducing a matrix angular central Gaussian (MACG) distribution over $k$-dimensional subspaces.
This distribution can be sampled efficiently using a low-rank decomposition of $\mathbf{S}$.
Let $\mathbf{S} = \mathbf{V}_r \diag(\boldsymbol{\lambda}_r) \mathbf{V}_r^\intercal$
be a compact EVD, where $r = \text{rank}(\mathbf{S})$.
The stochastic subspace can be written as:
\begin{equation}\label{eq:MACG-nkp-low-rank}
  \mathscr{W} = \mathbf{V}_r\, \pi_k\left(\diag(\boldsymbol{\lambda}_r)^{1/2}\,
    \mathbf{Z}_{r \times \beta}\right),
\end{equation}
where $\mathbf{Z}_{r \times \beta}$ is an $r$-by-$\beta$ standard Gaussian matrix.
This low-rank form allows SS-PPCA sampling to be performed in $O(r k \beta)$ flops, the same computational complexity as SS-Bootstrap.

However, because SS-PPCA is built on a Gaussian latent variable model, the resulting stochastic subspaces may struggle to capture non-Gaussian structure in the underlying data, such as skewness, heavy tails, or multimodality, potentially impacting its performance in such cases.

\subsection{Comparison of model structures} \label{sec:compare-model-structures}

To deepen understanding of the two methods, we outline their structures in \cref{fig:comparision}
and highlight the key differences below.
Both SS-PPCA and SS-Bootstrap (i) model principal subspaces, (ii) do not require discretization information, (iii) use a single scalar hyperparameter, and (iv) allow quantities of interest in the objective in training the hyperparameter.
Their primary distinction lies in modeling assumptions:
SS-PPCA posits a Gaussian latent-variable model on the system state and derives distributions for random matrices and their principal subspaces,
whereas SS-Bootstrap draws directly from the empirical data distribution via resampling to induce a distribution over principal subspaces.
The nonparametric nature of SS-Bootstrap allows it to adapt to the data distribution,
providing robustness to distributional misspecification.
By generating data-consistent stochastic subspaces, SS-Bootstrap leads to
SROMs that perturb the reference physics only along directions supported by the data,
allowing sharper, more reliable characterization of model error.

\subsection{Metrics for comparing stochastic models} \label{sec:metrics}

We evaluate the two stochastic subspace models using two general metrics: consistency and sharpness.
By \textit{consistent} UQ,
we mean that the statistics on prediction uncertainty---such as standard deviation
and predictive intervals (PI)---derived from the model match the prediction error on average.
Without consistency, the model either under- or over-estimates its prediction error,
which risks being over-confident or unnecessarily conservative. %
By \textit{sharp} UQ, we mean the model's PIs are narrower than alternative methods.
This is essentially the probabilistic version of model accuracy
and is desirable as we want to minimize error.

\section{Numerical examples}\label{sec:examples}
This section compares the proposed SS-Bootstrap method with SS-PPCA for characterizing model error using two numerical examples. 
First, we consider a parametric nonlinear static problem with non-Gaussian data.
Then, we consider a linear static problem to characterize model error using experimental data.
Lastly, we examine a linear dynamics problem involving a space structure. 
All data and code are available at \href{https://github.com/UQUH/SS_Bootstrap}{https://github.com/UQUH/SS_Bootstrap}.

\subsection{Parametric nonlinear static problem} \label{sec:param-static}

\begin{figure}[!t]
    \centering
    \includegraphics[width=0.9\textwidth]{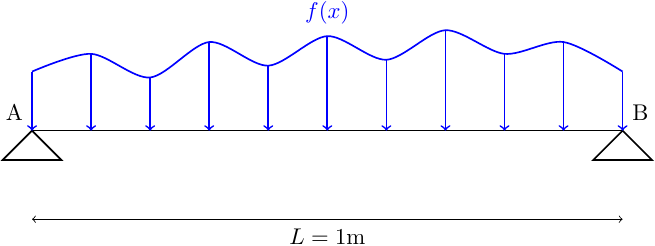}
    \caption{One-dimensional fixed-fixed nonlinear static system.}
    \label{fig:diagram-static}
\end{figure}

We consider a one-dimensional parametric nonlinear static problem (see \cref{fig:diagram-static})
with $n = \num{1000}$ degrees of freedom (DoFs), governed by
\begin{equation}
  \mathbf{K} \mathbf{u}(\boldsymbol{\mu}) + \alpha \mathbf{u}^3(\boldsymbol{\mu}) = \mathbf{f}(\boldsymbol{\mu}),
\end{equation}
where $\mathbf{u}(\boldsymbol{\mu}) \in \mathbb{R}^n$ is the displacement vector,
and $\boldsymbol{\mu} \in [0,1]^2$ is a parameter vector which controls the loading conditions.
The parameter $\alpha > 0$ controls the strength of the cubic nonlinearity. In this example, we set $\alpha = 10^4$.
Homogeneous Dirichlet boundary conditions are imposed at the first and last nodes, $x_1 (\boldsymbol{\mu}) = x_n (\boldsymbol{\mu})= 0$, which can be expressed compactly as $\mathbf{B}^\intercal \mathbf{x}(\boldsymbol{\mu}) = 0$ with $\mathbf{B} = [\mathbf{e}_1, \mathbf{e}_n]$, where $\mathbf{e}_i$ denotes the $i$-th standard basis vector.
The stiffness matrix $\mathbf{K} \in \mathbb{R}^{n \times n}$ is constructed as
$\mathbf{K}$ = $\mathbf{\Phi} \mathbf{\Lambda} \mathbf{\Phi}^{\intercal}$, with
$\mathbf{\Lambda} = \diag(4 \pi^2 j^2)_{j=1}^{n-2}$ and
$\mathbf{\Phi} = [0 \; \mathbf{S} \; 0]^\intercal$.
The matrix $\mathbf{S}$ = $\sqrt{\frac{2}{n-1}} \left[\sin \left(\frac{jk \pi}{n-1}\right)\right]_{k=1, \cdots, n-2}^{j=1, \cdots, n-2}$
is the order-$(n-2)$ type-I discrete sine transform (DST-I) matrix, scaled to be orthogonal.
Thus, $\mathbf{K}$ has eigenpairs $(\lambda_j, \boldsymbol{\phi}_j)$ with $\lambda_j = 4\pi^2 j^2$ and $\boldsymbol{\phi}_j$ is the $j$-th column of $\mathbf{\Phi}$.
The loading is parameterized as
\begin{equation}
    \mathbf{f}(\boldsymbol{\mu}) = \frac{\mathbf{g}(\boldsymbol{\mu})}{\|\mathbf{g}(\boldsymbol{\mu})\|_{\infty}}, \quad
    \mathbf{g}(\boldsymbol{\mu}) = \mu_1 (\boldsymbol{\phi}_2 + \boldsymbol{\phi}_3) + \mu_2(\boldsymbol{\phi}_4 + \boldsymbol{\phi}_5),
\end{equation}
with $\mu_1, \mu_2 \sim \text{Beta}(0.5,0.5),$
i.e., the force depends upon the parameters $\mu_i$.
This setup provides a controlled framework for studying parameter-dependent structural responses,
especially in the context of model error.

\begin{figure}[!t]
  \centering
    \centering
    \includegraphics[width=\textwidth]{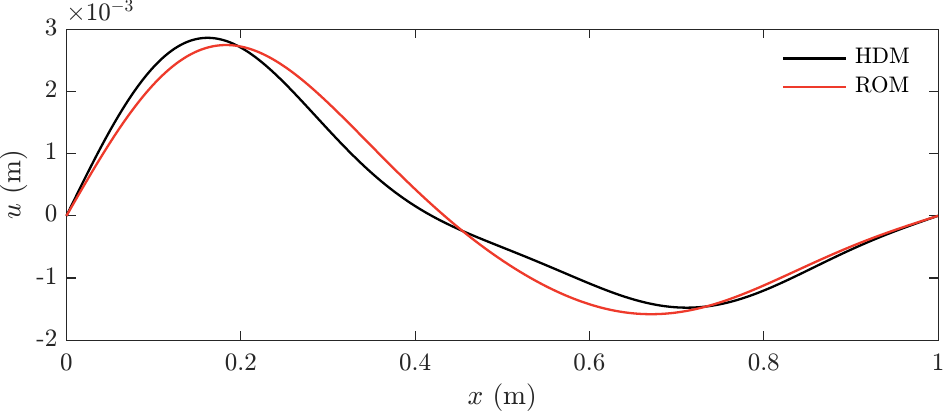}
    \caption{FOM vs ROM displacement at the test parameter}
    \label{fig:linear-static-HDM-ROM}
\end{figure}

The nonlinear FOM can be computationally expensive,
especially if it needs to be evaluated multiple times.
This justifies the use of ROM.
We construct a ROM by sampling 50 parameter points from the Beta-distribution, $\mu_1, \mu_2 \sim \text{Beta}(0.5,0.5)$.
For each sample, we solve $\mathbf{u}_j = \mathbf{K}^{-1} \mathbf{f}(\boldsymbol{\mu}_j)$, $j = 1, \cdots, 50$, and form the snapshot matrix $\mathbf{X} = [\mathbf{u}_1, \, \cdots, \, \mathbf{u}_{50}]$. 
The deterministic ROB $\mathbf{V} = \pi_k(\mathbf{X})$ with dimension $k=1$ is obtained via POD. 
This choice of $k=1$ is deliberate to produce a challenging ROM with a visible approximation error relative to the FOM, thereby providing a meaningful setting to evaluate the proposed stochastic subspace model's ability to characterize model error. If one chooses higher $k$ values, then it would enrich the deterministic POD space, improve the ROM approximation, and reduce the discrepancy that the SROM must represent.
The governing equation of the ROM is given by:
\begin{equation}
  \mathbf{u}_{R}(\boldsymbol{\mu}) = \mathbf{V}\mathbf{q}(\boldsymbol{\mu}),
    \quad \mathbf{V}^{\intercal} \mathbf{K} \mathbf{V} \, \mathbf{q}(\boldsymbol{\mu}) + \alpha \, \mathbf{V}^\top \left( \left( \mathbf{V} \, \mathbf{q}(\boldsymbol{\mu}) \right)^{3} \right) = \mathbf{V}^\top \mathbf{f}(\boldsymbol{\mu})
\end{equation}

To evaluate ROM accuracy, we define a test parameter
$\boldsymbol{\mu}_{\text{test}} = [\frac{1}{2},\frac{1}{2}]$,
which lies within the parameter domain but is not included in the training set.
\Cref{fig:linear-static-HDM-ROM} compares the FOM and ROM displacement at the test parameter.

To characterize ROM-induced error,
we construct SROMs by replacing the deterministic ROB $\mathbf{V}$ with SROBs $\mathbf{W}$.
For SS-Bootstrap, 
SROBs $\mathbf{W}$ are sampled using the stochastic subspace model
described in \cref{sec:stochastic-subspace-bootstrap} and \cref{alg:ss-Bootstrap}.
For SS-PPCA, 
SROBs $\mathbf{W}$ are sampled using the stochastic subspace model
described in \citep{yadav2025ss}.
In both cases, the hyperparameter $\beta$ is optimized by minimizing the objective function in \cref{eq:objective}, where $\mathbf{u}_E$ denotes FOM displacements at the training parameters and $\mathbf{u}_L^o$ denotes the corresponding ROM displacements.
For this example, the optimal $\beta$ obtained via SS-Bootstrap is 6, and the training process takes 62.3 seconds. Because the problem is nonlinear, the training time is slightly longer; however, this can be reduced by employing hyper-reduction techniques for nonlinear model order reduction \cite{Chaturantabut2010,Carlberg2010,Farhat2015}.

\begin{figure}[!t]
    \centering
    \includegraphics[width=0.95\linewidth]{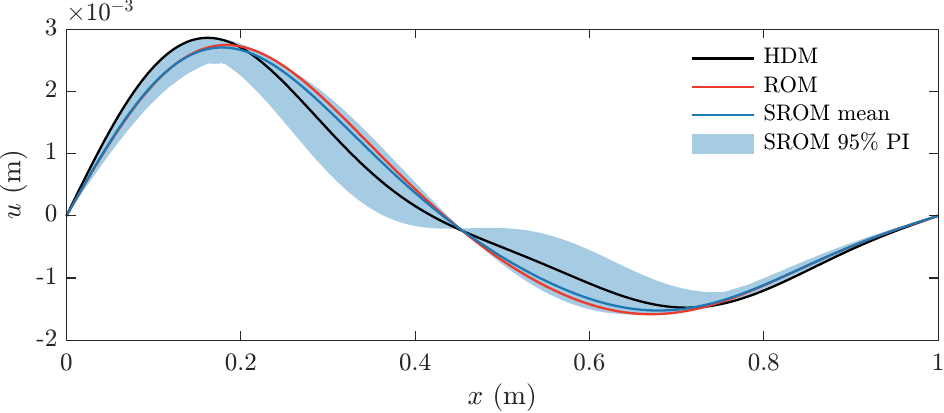}
    \caption{Nonlinear Static Problem: prediction using Bootstrap.}
    \label{fig:result-linear-static-Bootstrap}
\end{figure}

\begin{figure}[!t]
    \centering
    \includegraphics[width=0.95\linewidth]{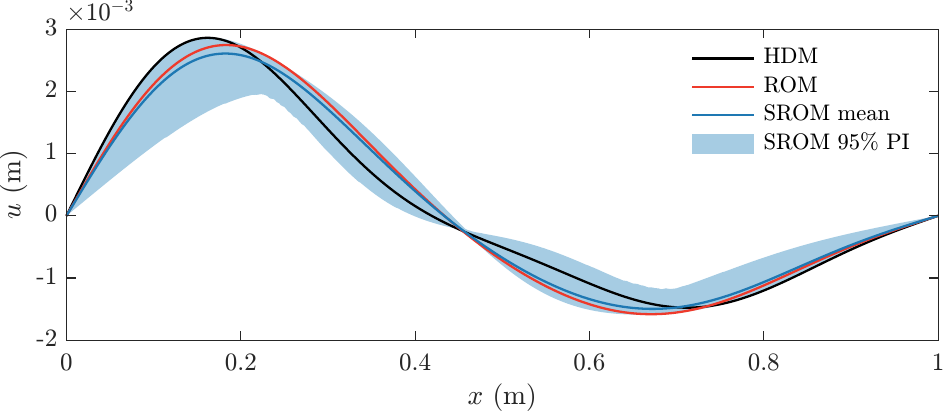}
    \caption{Nonlinear Static Problem: prediction using PPCA.}
    \label{fig:result-linear-static-PPCA}
\end{figure}

\Cref{fig:result-linear-static-Bootstrap,fig:result-linear-static-PPCA} compare SROM predictions from SS-Bootstrap and SS-PPCA for the nonlinear static example.
For both methods, the 95\% prediction interval (PI) is estimated using $\num{1000}$ Monte Carlo samples of the SROM, ensuring a consistent basis for comparison.
The empirical coverage—95.3\% for SS-Bootstrap and 94.9\% for SS-PPCA—indicates that both approaches produce uncertainty bounds that are statistically consistent with the underlying FOM response.
However, the average PI width for SS-PPCA is \(1.5\times\) that of SS-Bootstrap, highlighting the sharper uncertainty quantification provided by the proposed method.
This behavior reflects the underlying assumptions of the two methods.
SS-PPCA relies on a Gaussian latent-variable model, which can induce conservative subspace variability when the snapshot distribution is non-Gaussian, as is the case here due to the Beta-distributed loading parameters.
In contrast, SS-Bootstrap constructs stochastic subspaces directly from the empirical snapshot distribution, leading to subspace variability that is better aligned with the observed data and, consequently, to sharper PIs with comparable coverage.

\subsection{Static problem: characterizing FOM error} \label{sec:static}

In the previous example, we used SROM to characterize the truncation error induced by a ROM in a nonlinear parametric system.
However, the utility of SROM is not limited to ROM error characterization.
It can also be used to quantify discrepancies between a computational FOM and experimental data, which is central to model validation. To illustrate this setting, we consider here a linear static problem in which SROM is used to characterize FOM error from sparse experimental observations.

The synthetic experimental data is generated by solving the system:
\begin{equation}
  \mathbf{K}_{\text{E}} \mathbf{u}_{\text{E}} = \mathbf{f},
\end{equation}
with $n$ (model dimension) $= \num{1000}$,
$\mathbf{f} = \|\textbf{g}\|_{\infty}^{-1} \textbf{g}$
where $\textbf{g} = \sum_{i = 2}^{5} 0.5 \boldsymbol{\phi}^i + 1\boldsymbol{\phi}^6$
and homogeneous Dirichlet boundary conditions $\mathbf{B}^\intercal \mathbf{x}_{\text{E}} = 0$
where $\mathbf{B} = [\mathbf{e}_1 \, \mathbf{e}_n]$.
The stiffness matrix is defined as $\mathbf{K}_{\text{E}} = \mathbf{K} + \mathbf{K}_{\epsilon}$,
where $\mathbf{K}$ is same as the previous example
and $\mathbf{K}_{\epsilon}$ is
designed to induce a 15\% change in the Frobenius norm of $\mathbf{K}$.
In particular, 
$\mathbf{K}_{\epsilon}= \boldsymbol{\Phi}\boldsymbol{\Lambda}_{\epsilon} \boldsymbol{\Phi}^\intercal$ with $\boldsymbol{\Lambda}_{\epsilon} = \diag (c z_j \lambda_j)_{j=1}^{n-2}$, $z_j \overset{iid}{\sim} N(0,1)$, and $c$ is chosen such that $\| \boldsymbol{\Lambda}_{\epsilon}\|_{\text{F}} = 0.15  \|\boldsymbol{\Lambda}\|_{\text{F}}$. 
In addition to model discrepancy, measurement noise is introduced through
$\mathbf{u}_{\text{obs}}=\mathbf{u}_{\text{E}}+\boldsymbol{\eta}$, where $ \boldsymbol{\eta}\sim \mathcal{N}(0,\sigma^2\mathbf{I})$
with $\sigma^2$ is chosen to produce a $5\%$ noise level. Since experimental measurements are typically sparse, we assume that observations are available only at $19$ equidistant interior locations in the interval $(0,1)$. These sparse data are used to characterize FOM error through the SROM framework.

\begin{figure}[!t]
  \centering
  \includegraphics[width=\textwidth]{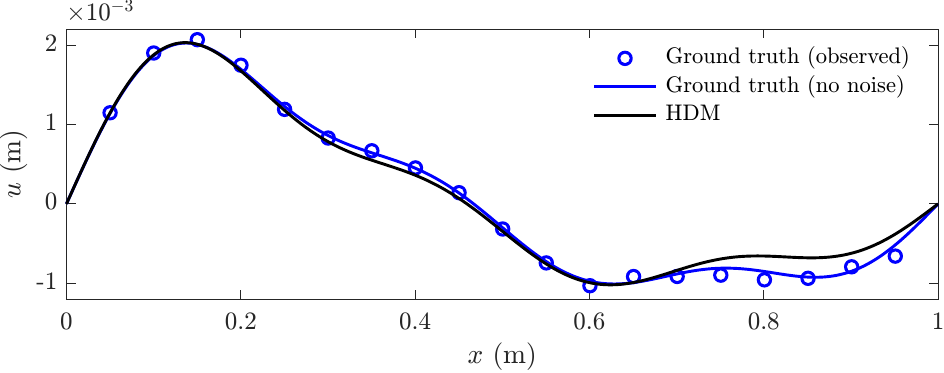}
  \caption{Ground truth vs FOM displacement.}
  \label{fig:ex3-HDM-EXP}
\end{figure}

The computational FOM is defined as:
\begin{equation}
  \mathbf{K} \mathbf{u}_{\text{H}} = \mathbf{f},
\end{equation}
with the same $\mathbf{K}$ and $\mathbf{f}$ as above.
However, the FOM produces biased predictions because the true system differs from this model, see \cref{fig:ex3-HDM-EXP}. To capture this error using SROM,
we collect snapshots with perturbed FOMs
so that the perturbed FOM responses can sufficiently cover the experimental data.
The stiffness matrix of the FOM is perturbed 100 times in a way similar to the construction of
$\mathbf{K}_\text{E}$, and the responses are collected in a snapshot matrix.
ROM of dimension $k = 4$ is then constructed using POD, which can be written as:
\begin{equation}
  \mathbf{u}_{\text{R}} = \mathbf{V}\mathbf{q},
  \quad \mathbf{V}^{\intercal} \mathbf{K} \mathbf{V} \mathbf{q} = \mathbf{V}^{\intercal}\mathbf{f}.
\end{equation}
The corresponding SROM is obtained by replacing the deterministic ROB $\mathbf{V}$ with a stochastic ROB $\mathbf{W}$. The basis $\mathbf{W}$ is sampled from the stochastic subspace model described in \cref{sec:stochastic-subspace-bootstrap} and \cref{alg:ss-Bootstrap}. The hyperparameter $\beta$ is trained using the criterion in \cref{eq:objective} together with the sparse observed data. In this example, $\mathbf{u}_{\text{E}}^{o}$ denotes the sparse experimental displacement observations, and $\mathbf{u}_{\text{R}}^{o}$ denotes the ROM displacement evaluated at the same locations. Using the training strategy in \cref{sec:training}, the optimal value of $\beta$ for SS-Bootstrap is found to be $20$.

\begin{figure}[!t]
  \centering
  \includegraphics[width=\textwidth]{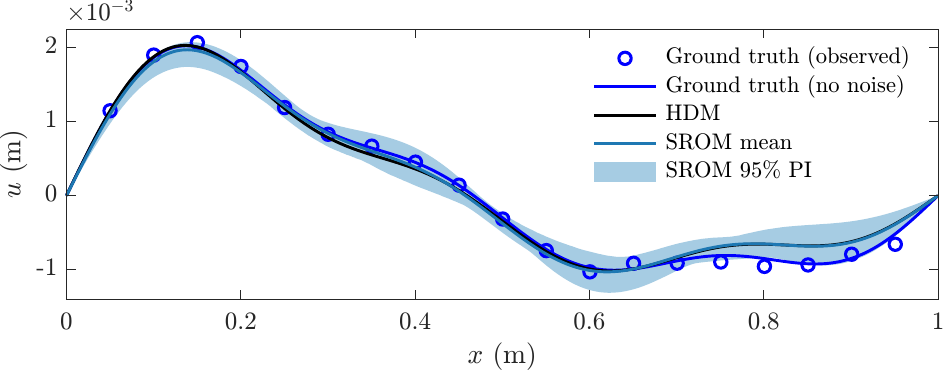}
  \caption{Linear static problem with experimental data: SS-Bootstrap prediction.}
  \label{fig:ex2-bootstrap-prediction}
\end{figure}

\begin{figure}[!t]
  \centering
  \includegraphics[width=\textwidth]{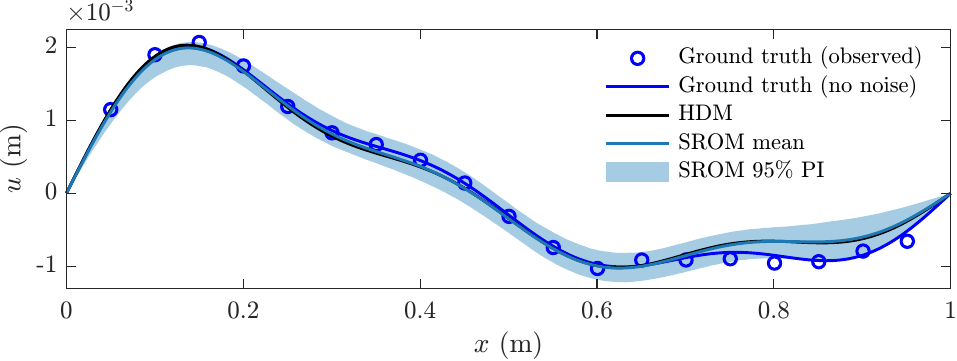}
  \caption{Linear static problem with experimental data: SS-PPCA prediction.}
  \label{fig:ex2-PPCA-prediction}
\end{figure}

\Cref{fig:ex2-bootstrap-prediction,fig:ex2-PPCA-prediction} compare the SROM predictions obtained from SS-Bootstrap and SS-PPCA. The empirical coverage is $95.0\%$ for SS-Bootstrap and $94.2\%$ for SS-PPCA, indicating that both methods produce statistically consistent uncertainty bounds for the underlying ground-truth displacement. In this example, the average prediction-interval width for SS-PPCA is approximately $1.05$ times that of SS-Bootstrap, demonstrating the consistent uncertainty quantification provided by the proposed method.

\subsection{Linear dynamics: space structure}\label{sec:dynamic}

\begin{figure}[!t]
    \centering
    \includegraphics[width=\linewidth]{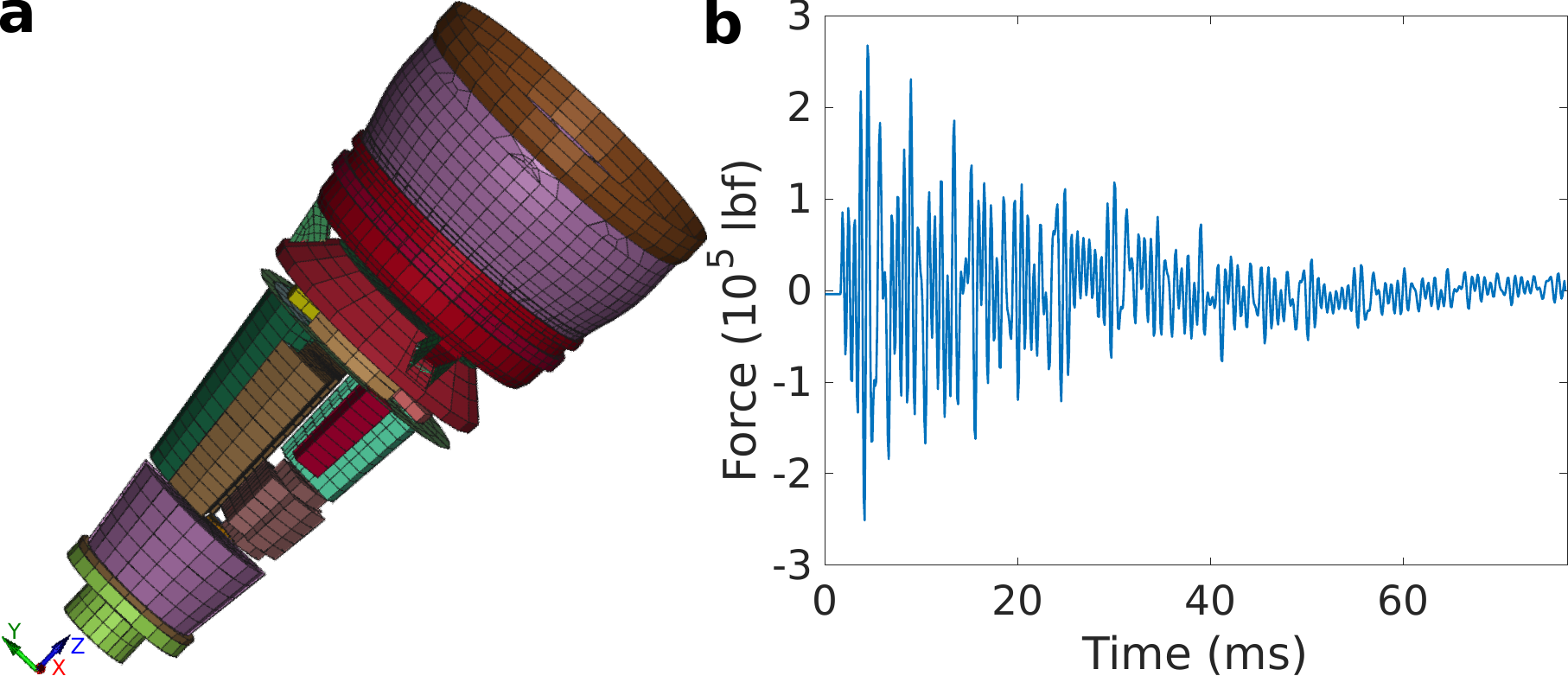}
    \caption{Linear Dynamics: space structure.}
    \label{fig:space-structure}
\end{figure}

We consider a major component of a space structure \citep{ZengX2023},
see \cref{fig:space-structure}a.
The space structure comprises two parts:
an open upper section and a lower section.
A large mass (approximately 100 times heavier than the remainder of the model)
is located at the center of the upper section and is connected to the sidewalls via rigid links.
The lower section consists of an outer cylindrical shell
and an inner shock absorption block with a hollow cavity that houses essential components.
The finite element model includes both solid and shell elements with linear-elastic material behavior. The outer part is modeled using shell elements, whereas the shock-absorption blocks in the lower section are modeled using solid elements. The structure has a maximum dimension of approximately 8 meters in both the 
$x$- and $y$-directions and a height of 17.3 meters.
The impulse load (\cref{fig:space-structure}b)
is applied at the mass center along the $z$-direction, transmitted to the upper section through the sidewalls,
and then to the lower section via the mounting pedestal.
Because the load can compromise the functionality of essential components, we monitor acceleration and velocity at critical points to ensure safe operation.
The quantity of interest (QoI) is the $x$-velocity at a critical point on one essential component.
The space structure is modeled as a free-free structure, i.e., no prescribed displacement or support boundary conditions are imposed. Accordingly, the exported LS-DYNA model is used in a transient dynamic setting, for which rigid-body motion is admissible.
The governing equation of the FOM of the space structure is given by:
\begin{equation}\label{eq:HDM-space}
  \mathbf{M}_{\text{H}} \, \ddot{\mathbf{x}} + \mathbf{C}_{\text{H}} \, \dot{\mathbf{x}} + \mathbf{K}_{\text{H}} \, \mathbf{x} =
  \mathbf{f},
\end{equation}
with initial conditions $\mathbf{x}(0)=\mathbf{0}$ and $\dot{\mathbf{x}}(0)=\mathbf{0}$. The matrices $\mathbf{M}_{\text{H}}$, $\mathbf{K}_{\text{H}}$, and the force vector $\mathbf{f}$ are exported from an LS-DYNA finite element model; the damping is Rayleigh, $\mathbf{C}_{\text{H}}=\beta_{\text{H}}\mathbf{K}_{\text{H}}$, with $\beta_{\text{H}}=\num{6.366e-6}$. The FOM response $\mathbf{x}_{\text{H}}$ is computed via the Newmark-$\beta$ method with a time step of $5\times 10^{-2}$ ms. One FOM run takes approximately 38 minutes, which is prohibitive for repeated evaluations.
To reduce cost, we construct a POD-based ROM of dimension $k=10$.
The ROM derived from the FOM is defined as
\begin{equation}
  \mathbf{x}_{\text{R}} = \mathbf{V}\mathbf{q},
  \quad \mathbf{M}_{\text{H},\mathbf{V}} \, \ddot{\mathbf{q}} + \mathbf{C}_{\text{H},\mathbf{V}} \, \dot{\mathbf{q}} + \mathbf{K}_{\text{H},\mathbf{V}} \, \mathbf{q} = \mathbf{V}^{\intercal} \mathbf{f},
\end{equation}
where reduced matrices are denoted with a pattern $\mathbf{A}_{\mathbf{V}} := \mathbf{V}^{\intercal} \mathbf{A} \mathbf{V}$,
so that $\mathbf{M}_{\text{H},\mathbf{V}} = \mathbf{V}^{\intercal} \mathbf{M}_{\text{H}} 
\mathbf{V}$, for example.
The initial condition of ROM is given by $\mathbf{q}(0) = 0$ and $\dot{\mathbf{q}}(0) = 0$.
The ROM runs in $0.2$ seconds—about $1.12\times 10^{4}$ times faster than the FOM—at the expense of approximation error.
We therefore use SROMs for error characterization by replacing the deterministic ROB $\mathbf{V}$ with stochastic ROBs (SROBs) $\mathbf{W}$.

For SS-Bootstrap, $\mathbf{W}$ is sampled using the stochastic subspace model defined in \cref{sec:stochastic-subspace-bootstrap} and \cref{alg:ss-Bootstrap}.
For SS-PPCA, $\mathbf{W}$ is sampled using the probabilistic PCA–based model described in \citep{yadav2025ss}.
Here, $\mathbf{u}_E$ is the FOM velocity at the critical node and $\mathbf{u}_L^o$ is the ROM velocity at the same location.
The optimal hyperparameter via the bootstrap method is $\beta=69$, obtained using the procedure in \Cref{sec:training}. Training takes $19.4$ seconds, which is negligible relative to an FOM run, indicating good scalability.

\begin{figure}[!tbp]
    \centering
    \includegraphics[width=0.7\linewidth]{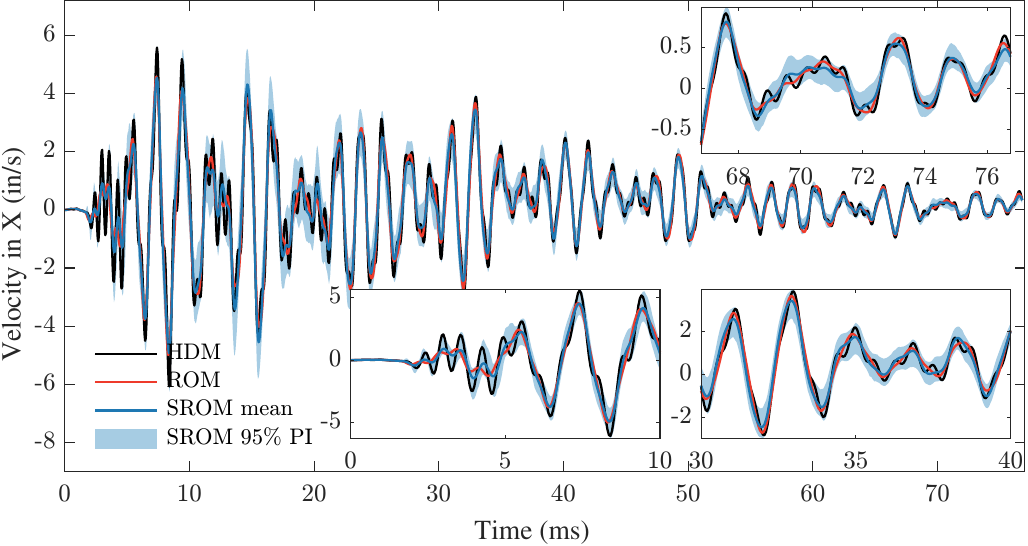}
    \caption{Dynamic prediction of the Bootstrap model: velocity.}
    \label{fig:prediction-Bootstrap-vel}
\end{figure}

\begin{figure}[!tbp]
    \centering
    \includegraphics[width=0.7\linewidth]{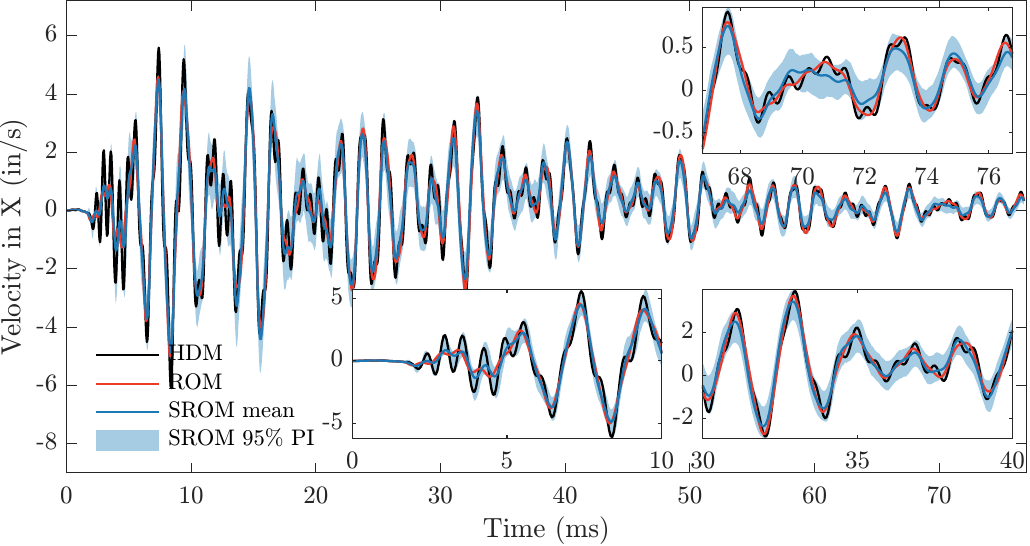}
    \caption{Dynamic prediction of the PPCA model: velocity.}
    \label{fig:prediction-PPCA-vel}
\end{figure}

\Cref{fig:prediction-Bootstrap-vel,fig:prediction-PPCA-vel} compare the velocity prediction of SS-Bootstrap and SS-PPCA.
Although the two methods yield visually similar envelopes, SS-Bootstrap consistently produces sharper PIs (see \cref{tab:dynamics}).
The FOM velocity exhibits high-frequency behavior during the initial $0$–$10$ ms, which ROM cannot fully capture.
Both stochastic models capture this error well, with SS-PPCA achieving reasonable coverage (93.34\%) but wider PIs, whereas SS-Bootstrap attains higher coverage (95.58\%) with tighter bounds.

\begin{table}[!tbp]
  \centering
  \begin{threeparttable}
    \caption{Dynamics problem: consistency and sharpness of 95\% PI.}
    \begin{tabular}{lllllc}
      \toprule
      & \multicolumn{2}{c}{SS-PPCA} & \multicolumn{2}{c}{SS-Bootstrap} & \\
      \cmidrule{2-5} %
      95\% PI & coverage\textsuperscript{*} & width\textsuperscript{\dag} & coverage\textsuperscript{*} & width\textsuperscript{\dag} & width ratio\textsuperscript{\ddag} \\
      \midrule
      $d_x$ ($10^{-4}$ in) & \underline{96.11\%} & 3.17 & \textbf{97.08\%} & \textbf{2.69} & \textbf{1.37} \\
      $v_x$ (in/s) & 93.34\% & 1.18 & \underline{\textbf{95.58\%}} & \textbf{1.11} & \textbf{1.19} \\
      $a_x$ ($10^3$ in/s\textsuperscript{2}) & 84.14\% & 7.26 & \underline{\textbf{87.06\%}} & \textbf{7.22} & \textbf{1.04} \\
      $v_r$ (in/s) & \underline{93.42}\% & \textbf{1.41} & \textbf{96.63\%} & 1.51 & \textbf{1.11} \\
      \bottomrule
    \end{tabular}
    \begin{tablenotes}
    \item[*] Proportion of true response covered by the 95\% PI: larger numbers are in boldface; numbers closer to 95\% are in underline.
    \item[\dag] Average width of of 95\% PIs: smaller numbers are in boldface.
    \item[\ddag] Average ratio of 95\% PI widths, PPCA vs. Bootstrap: numbers larger than one are in boldface.
    \end{tablenotes}
    \label{tab:dynamics}
  \end{threeparttable}
\end{table}


\begin{figure}[!tbp]
    \centering
    \includegraphics[width=0.7\linewidth]{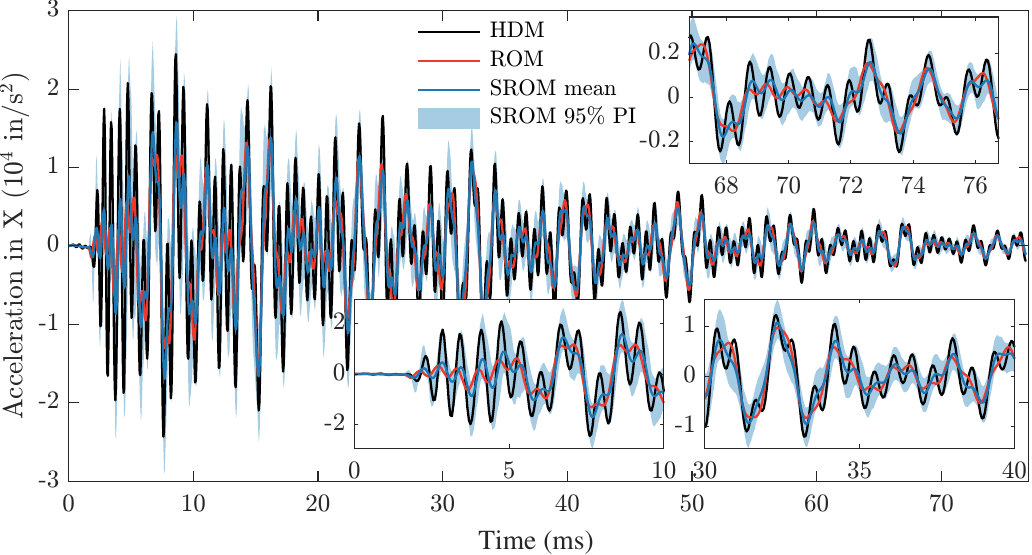}
    \caption{Dynamic prediction of the Bootstrap model: acceleration.}
    \label{fig:prediction-Bootstrap-acc}
\end{figure}

\begin{figure}[!tbp]
    \centering
    \includegraphics[width=0.7\linewidth]{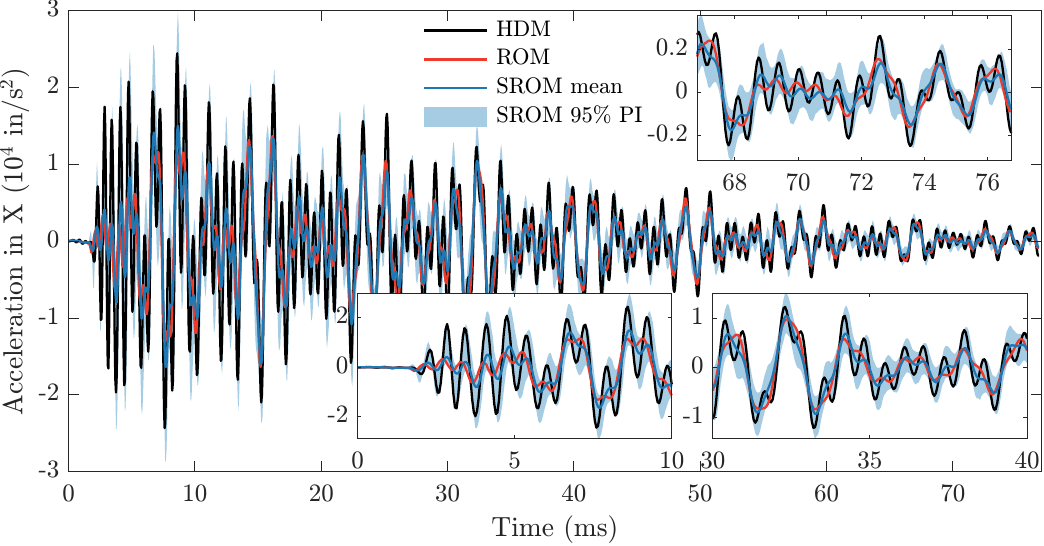}
    \caption{Dynamic prediction of the PPCA model: acceleration.}
    \label{fig:prediction-PPCA-acc}
\end{figure}

A similar trend appears for acceleration predictions (\cref{fig:prediction-Bootstrap-acc,fig:prediction-PPCA-acc}).
Both methods track the overall FOM response, but SS-Bootstrap again yields narrower PIs while maintaining higher coverage (87.06\% vs.\ 84.14\%).
This indicates that the bootstrap-based approach captures the high-frequency variability introduced by the impulse more effectively than SS-PPCA, benefiting from its direct use of the empirical snapshot distribution rather than a Gaussian latent-variable model.

It is important to note that, although the hyperparameter is trained using velocity (the QoI in this example) and the deterministic ROB \(\mathbf{V}\) is obtained from displacement snapshots, the approach enables predictions for unobserved quantities such as acceleration. More broadly, it supports the efficient prediction of unobserved QoIs across the system—a capability often lacking in related studies—thereby enhancing its practical applicability.

\begin{figure}[!tbp]
    \centering
    \includegraphics[width=0.7\linewidth]{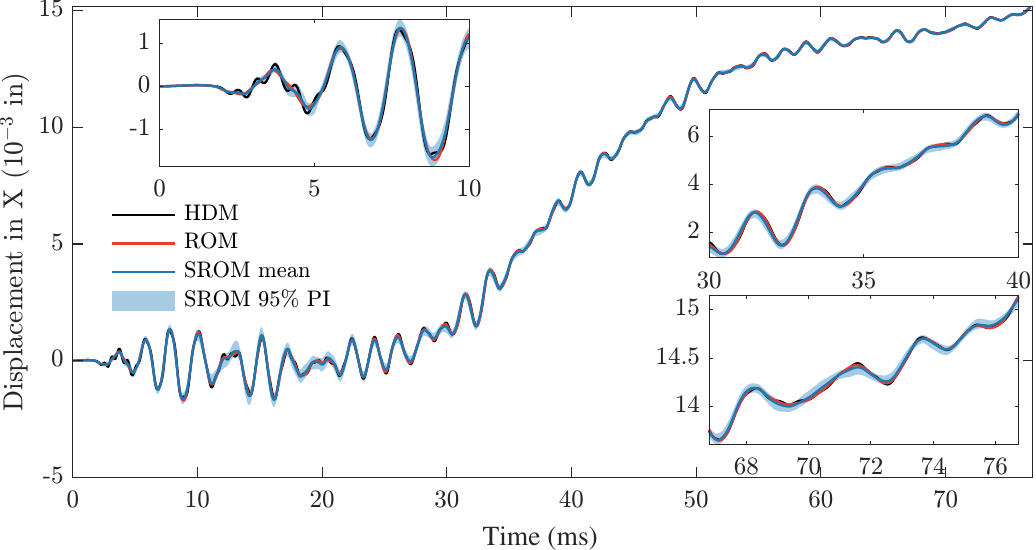}
    \caption{Dynamic prediction of the Bootstrap model: displacement.}
    \label{fig:prediction-Bootstrap-disp}
\end{figure}

\begin{figure}[!tbp]
    \centering
    \includegraphics[width=0.7\linewidth]{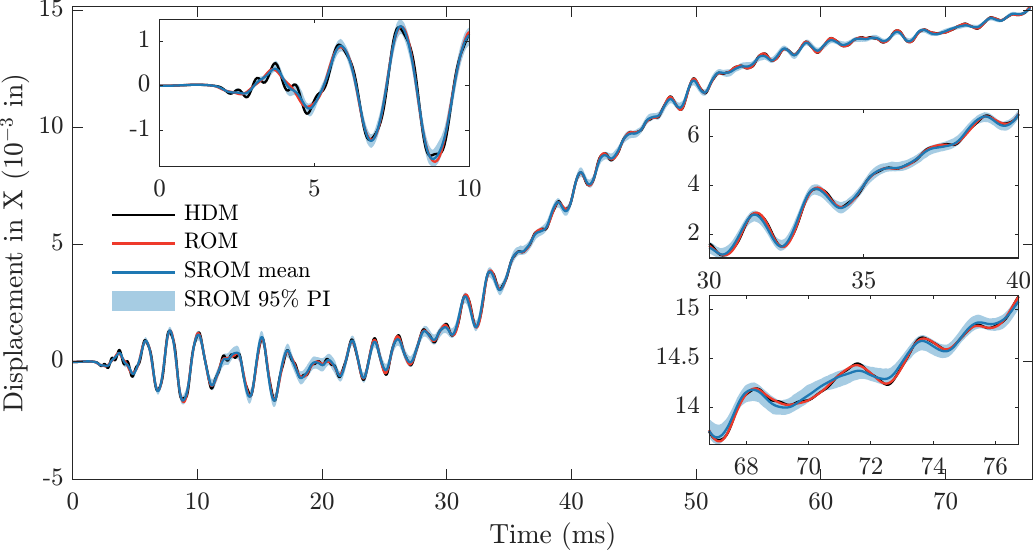}
    \caption{Dynamic prediction of the PPCA model: displacement.}
    \label{fig:prediction-PPCA-disp}
\end{figure}

\Cref{fig:prediction-Bootstrap-disp,fig:prediction-PPCA-disp} report the displacement predictions at the same node.
SS-PPCA provides accurate and consistent estimates (96.11\% coverage), yet SS-Bootstrap achieves comparable or better coverage (97.08\%) with substantially narrower PIs (see \cref{tab:dynamics}), demonstrating improved sharpness across QoIs.

\begin{figure}[!tbp]
    \centering
    \includegraphics[width=0.7\linewidth]{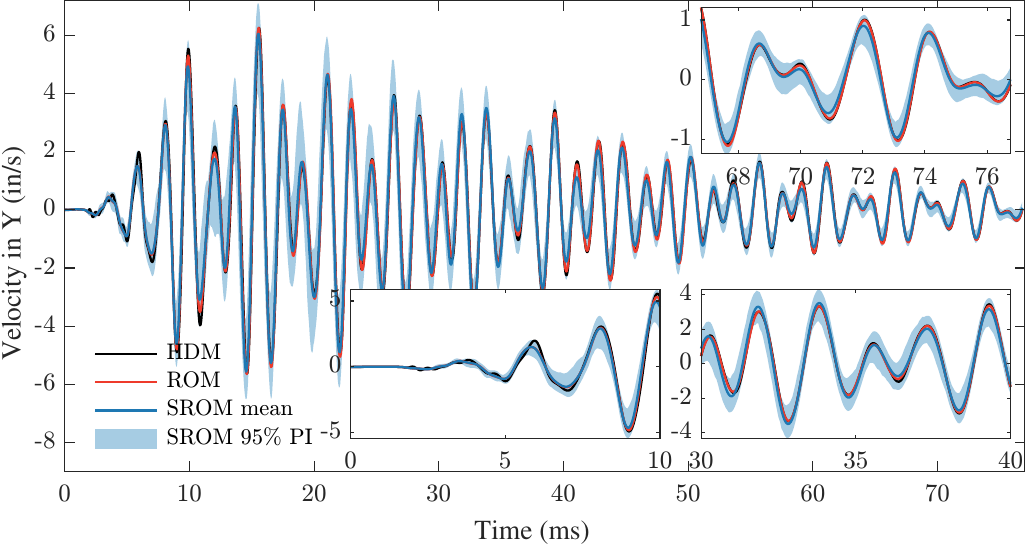}
    \caption{Dynamic prediction of the Bootstrap model: velocity at a random dof.}
    \label{fig:prediction-Bootstrap-vel-rand}
\end{figure}

\begin{figure}[!tbp]
    \centering
    \includegraphics[width=0.7\linewidth]{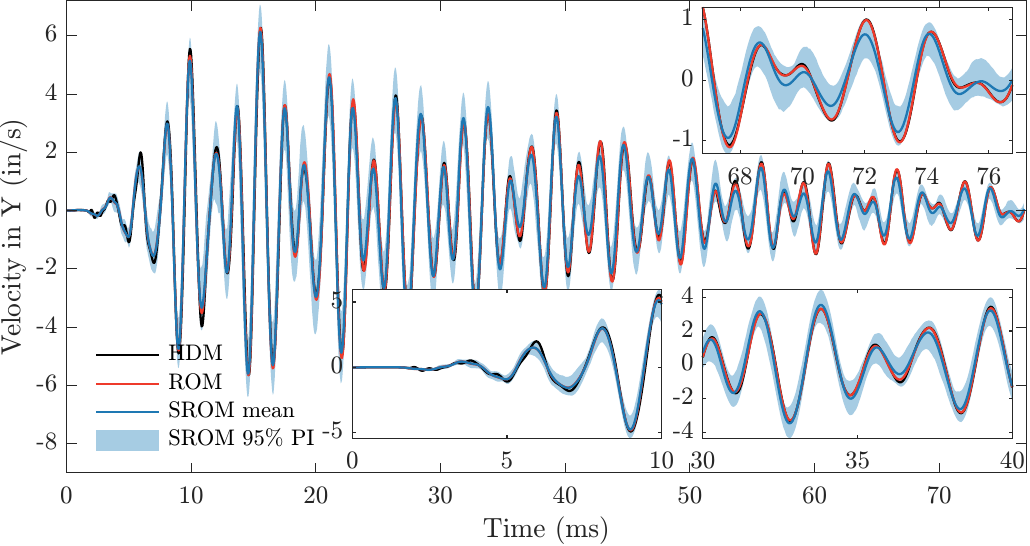}
    \caption{Dynamic prediction of the SS-PPCA model: velocity at a random dof.}
    \label{fig:prediction-PPCA-vel-rand}
\end{figure}

While the results above focus on a single critical node, many applications require predictions at other locations or system-level behavior.
Both SROM approaches generalize to unobserved DOFs, as shown in \cref{fig:prediction-Bootstrap-vel-rand,fig:prediction-PPCA-vel-rand}. 
The SROM mean under SS-Bootstrap remains closer to both the FOM and deterministic ROM predictions, whereas SS-PPCA exhibits slightly larger deviations. 
Nonetheless, both methods maintain consistency, with SS-Bootstrap providing uniformly narrower PIs and thus more efficient uncertainty quantification for system-level predictions.

SS-PPCA and SS-Bootstrap provide automated and efficient training with a single hyperparameter. While SROM-based methods cannot fully eliminate model error, since they are built on an underlying reduced-order model, their purpose is not to improve deterministic ROM accuracy directly. Rather, they provide a stochastic characterization of the residual error under a fixed reduced dimension, thereby enabling calibrated probabilistic predictions. In particular, each SROM realization remains $k$-dimensional; the improvement comes not from enlarging the dimension of an individual ROM, but from sampling a distribution of nearby $k$-dimensional subspaces instead of relying on a single deterministic POD subspace. This perspective is useful both when the residual error arises from model reduction and when the reduced model is used to characterize broader discrepancies between the computational model and the target ground truth. As $k$ increases, the deterministic ROM generally becomes more accurate, so the residual error requiring stochastic characterization is expected to decrease. In regimes where the ROM is already highly accurate, the additional benefit of stochastic subspace modeling may therefore become less pronounced. The accuracy of ROM can also be improved by adding closure terms; see \citet{Ahmed2021} for a review of ROM closure.

\section{Conclusion} \label{sec:conclusion}
We introduced a bootstrap-based stochastic subspace model for characterizing model error within the framework of stochastic reduced-order models.
The proposed SS-Bootstrap method is simple, efficient, and straightforward to implement.
It requires only one hyperparameter, which is optimized systematically to ensure consistency.

Through numerical examples, we demonstrated that SS-Bootstrap provides sharper and more reliable uncertainty quantification than the existing SS-PPCA approach.
The method consistently captured different sources of model error, including parametric uncertainty and ROM truncation error, while maintaining a low computational cost.

While the bootstrap procedure presented in this paper applies to constructing randomized
principal subspaces of data matrices and is applied to POD-based reduced-order models,
the same procedure may also be applied to other sampling-based model reduction approaches,
such as the reduced basis method and dynamic mode decomposition.
Moreover, the framework integrates naturally with methods for propagating aleatory uncertainty.
Each random realization of the aleatory inputs can be propagated through a random realization of the SROM, yielding an ensemble that captures the combined effects of all uncertainties in a consistent and computationally tractable way.

In regimes with very large errors, however, characterization alone may be insufficient.
Future work will therefore investigate practical strategies for correcting model errors, building on the robust characterization enabled by SS-Bootstrap.

\section{Data Availability Statement}

Some or all data, models, or code generated or used during the study are available in a repository online in accordance with funder data retention policies \href{https://github.com/UQUH/SS_Bootstrap}{https://github.com/UQUH/SS_Bootstrap}.

\section{Acknowledgments}
This work was funded by the University of Houston through the SEED program no. 000189862.

\bibliographystyle{ascelike-new}
\bibliography{Bootstrap}

\end{document}